\documentclass[epj]{webofc}
\usepackage[varg]{txfonts}   
\usepackage{lineno}
\begin{document}
%
\title{Ultra High Energy Cosmic Rays: \\
Origin, Composition and Spectrum}
%
%

\author{Roberto Aloisio\inst{1,2}\fnsep\thanks{\email{roberto.aloisio@gssi.it}}}

\institute{Gran Sasso Science Institute, L'Aquila, Italy
\and
    	INFN - Laboratori Nazionali del Gran Sasso, Assergi (L'Aquila), Italy. 
 }

\abstract{%
The physics of Ultra High Energy Cosmic Rays will be reviewed, discussing the latest experimental results and theoretical models aiming at explaining the observations in terms of spectra, mass composition and possible sources. It will be also discussed the emission of secondary particles such as neutrinos and gamma rays produced by the interaction of Ultra High Energy Cosmic Rays with astrophysical photon backgrounds. The content of the present proceeding paper is mainly based on the review papers \cite{Aloisio:2017qoo,Aloisio:2017ooo}.}
\maketitle
\section{Introduction}
\label{intro}

Ultra High Energy Cosmic Rays (UHECR) are the most energetic particles observed in nature, with energies exceeding $10^{20}$ eV. The study of these particles dates back to 1962 when J. Linsley in the Volcan Ranch experiment observed the first extensive air shower generated by a cosmic particle of estimated energy around $10^{20}$ eV (see \cite{Aloisio:2017qoo,Aloisio:2017ooo} and references therein).

Nowadays, the most advanced experiments to detect UHECR are the Pierre Auger Observatory in Argentina \cite{ThePierreAuger:2015rma}, far the largest experimental setup devoted to the study of UHECR, and the Telescope Array (TA) experiment \cite{Tinyakov:2014lla}, placed in the United States, with roughly $1/10$ of the Auger statistics. 

Both detectors exploit the hybrid concept, combining an array of surface detectors (SD) to sample Extensive Air Showers (EAS, shower) when they reach the ground and telescopes, overlooking the surface array, to collect the fluorescence light of the atmospheric nitrogen excited by the EAS (fluorescence detectors, FD). The advent of the hybrid approach has been a major breakthrough in the detection of UHECR since, combining the measurements of the SD and FD, the tiny flux of UHECRs is studied with a 100$\%$ duty cycle and with an almost calorimetric estimation of the shower energies.

The experimental study of UHECR clarified few important characteristics of these particles: (i) UHECR are charged particles with a limit on photon and neutrino fluxes around $10^{19}$ eV at the level of few percent and well below respectively (see \cite{Aloisio:2017qoo,Aloisio:2017ooo} and references therein), (ii) the spectrum observed at the Earth shows a slight flattening at energies around $5\times 10^{18}$ eV (called the ankle) with (iii) a steep suppression at the highest energies (see \cite{Aloisio:2017qoo,Aloisio:2017ooo} and references therein), (iv) the arrival directions of particles with $E>5\times 10^{18}$ eV show a strong (at the level of 5.4$\sigma$) dipole anisotropy  signalling the extragalactic origin of these particles \cite{Aab:2017tyv}.

Together with the number of particles per unit energy arriving at the Earth (flux) and their arrival direction (anisotropy) the third observable that characterises cosmic rays phenomenology is the mass composition. In the case of UHECR, mass composition is still matter of some debate. Before the advent of Auger the experimental evidences were all pointing toward a light composition with a proton dominated flux until the highest energies (see \cite{Aloisio:2017qoo,Aloisio:2017ooo} and references therein). The measurements carried out by the Auger observatory \cite{Aab:2014kda} have shown that the mass composition, from prevalently light at $\sim 10^{18}$ eV, becomes increasingly heavier towards higher energies. If confirmed, these findings would represent a change of paradigm respect to the picture of ten years ago. On the other hand, the TA experiment, even if with $1/10$ of the Auger statistics, collected data that seem to confirm the pre-Auger scenario \cite{Abbasi:2014sfa}, the mass composition is compatible with being light for energies above $10^{18}$ eV. Nevertheless, the analysis performed by the joint working group Auger-TA showed that the results of the two detectors are compatible within the systematic uncertainties \cite{TheTelescopeArray:2018dje}.

A clear experimental determination of mass composition is a difficult task as it is performed in an indirect way through the observation of the EAS development in the atmosphere and its comparison with the results of Monte Carlo simulations. These simulations are affected by the uncertainties in the hadronic interaction model between primary CR and the nuclei of the atmosphere. Direct measurements of the interaction cross sections are performed (at the LHC accelerator in CERN) only till (center of mass) energies around 10 TeV and therefore extrapolated at the interesting energies for UHECR (up to 500 TeV). As discussed by several authors (see \cite{Aloisio:2017qoo,Aloisio:2017ooo} and references therein), the cross section extrapolation, being model dependent, introduces a certain level of uncertainty, particularly at the highest energies, that could even spoil a clear determination of the mass composition itself. 

In order to interpret the observations at the Earth, trying to constrain the possible sources of UHECR, it is very important a detailed modelling of UHECR propagation in the intergalactic medium, which is mainly conditioned by the interaction with astrophysical photon backgrounds and intergalactic magnetic fields, being the presence of these fields still matter of debate we will not touch here the effect of magnetic fields on the propagation of UHECR. The interactions with astrophysical photon backgrounds (CMB, Cosmic Microwave Background, and EBL, Extragalactic Background Light) shape the spectrum of UHECR observed at the Earth and are also responsible for the production of secondary (cosmogenic) particles: photons and neutrinos. This secondary radiation can be observed through ground-based or satellite experiments and brings important informations about the mass composition of UHECR and, possibly, on their sources. 

The study of UHECR is a vital and intriguing field of research with a wide impact also in fundamental physics and cosmology, as it involves particles with extreme energies not attainable in any Earth based laboratory. This is the case of theories with Lorentz invariance violations or super heavy dark matter that in principle can be tested through UHECR observations, we will not discuss this issues here interested readers can refer to \cite{Aloisio:2017qoo,Aloisio:2015lva} and reference therein.

In the forthcoming sections we will briefly review the main theoretical models on UHECR sources compatible with observations (section \ref{sources}) and the production of secondary cosmogenic gamma rays and neutrinos (section \ref{secondaries}). We will conclude in section \ref{conclusions}.

\section{Source models}
\label{sources}

Following the discussion presented in \cite{Aloisio:2017qoo,Aloisio:2017ooo}, to constrain the basic characteristics of UHECR sources we adopt a purely phenomenological approach in which sources are homogeneously and isotropically distributed with the basic parameters: $\gamma_g$ injection power law index, $E_{max}$ maximum energy at the source (rigidity dependent), ${\cal L}_S$ emissivity and relative abundances of different elements at injection. These parameters are fitted to experimental data (both spectrum and mass composition) with as little as possible {\it a priori}  theoretical prejudice on what the values should be.

\begin{figure}[!h]
\centering
\includegraphics[scale=.25]{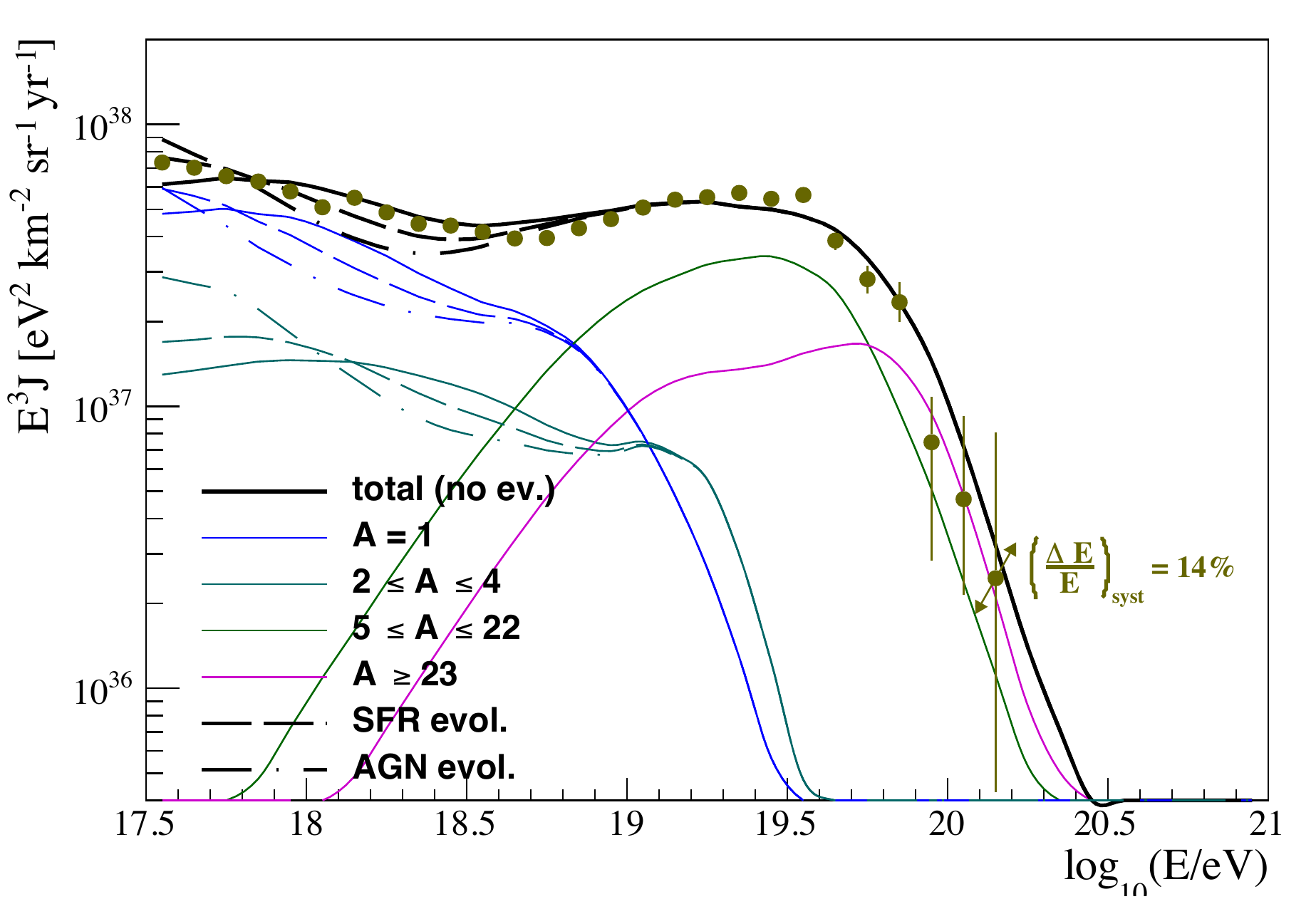}
\includegraphics[scale=.25]{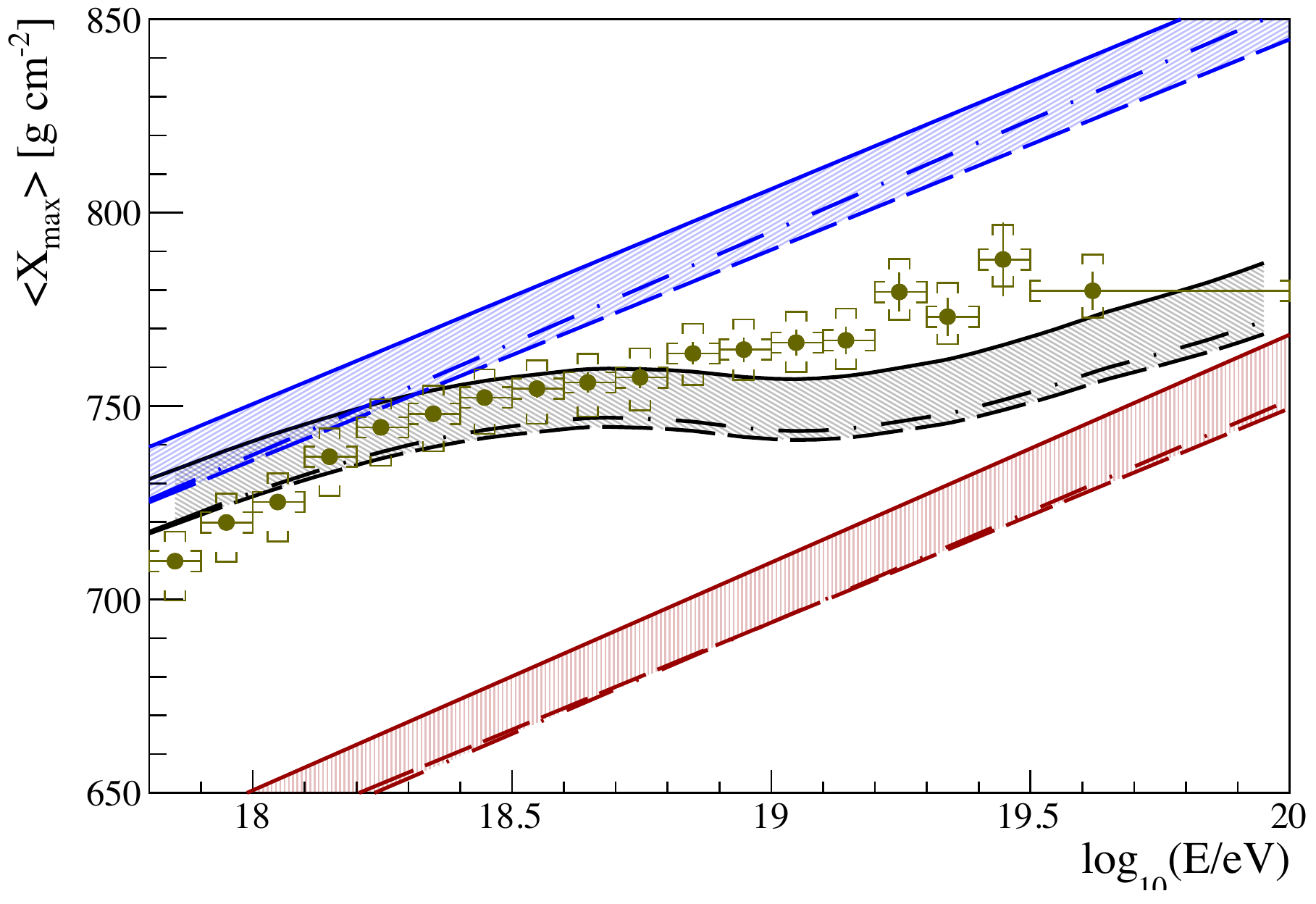}
\includegraphics[scale=.25]{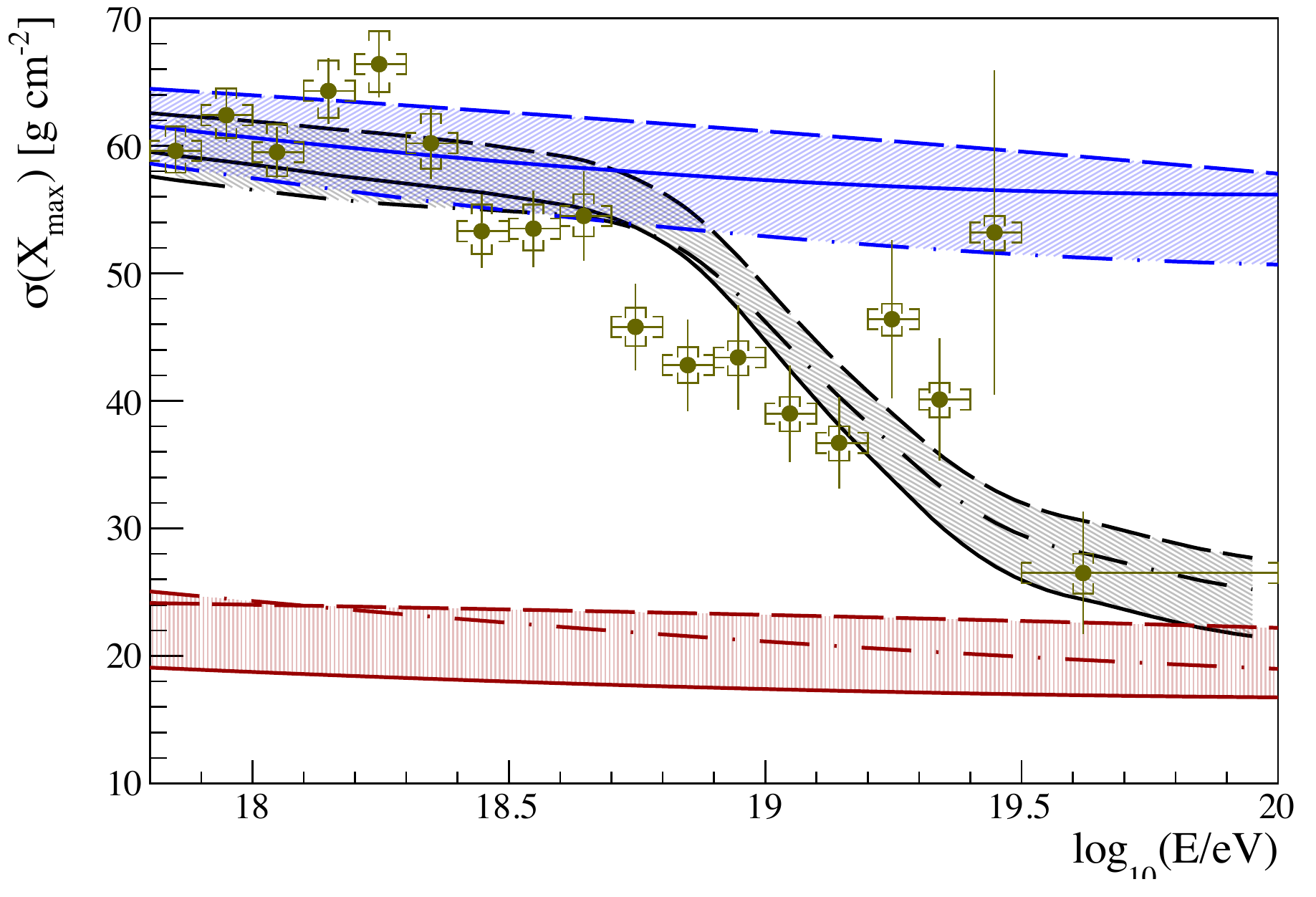}
\caption{Comparison of the flux (left panel), elongation rate (central panel) and its root mean square (right panel) as observed by Auger and computed assuming the model with two classes of different extragalactic sources \cite{Aloisio:2015ega}.}
\label{fig1}   
\end{figure}

The interactions suffered by UHECR with astrophysical photons backgrounds are pair-production, photo-pion production and, only for nuclei heavier than protons, photo-disintegration. Taking into account these channels of energy losses, together with the adiabatic energy losses due to the expansion of the universe, solving the transport equations for UHECR we can determine the theoretical flux, to be compared with observations. Mass composition is inferred from the mean value of the depth of shower maximum $\langle X_{max} \rangle$ and its dispersion (RMS) $\sigma(X_{max})$, computed as shown in \cite{Abreu:2013env} and compared with observations too. The combined analysis of $\langle X_{max} \rangle$ and $\sigma(X_{max})$, even if not conclusive allows to obtain less model dependent information on the mass composition. 

The qualitative new finding of Auger that mass composition might be mixed has served as a stimulus to build models that can potentially explain the phenomenology of Auger data. These models all show that the Auger spectrum and mass composition at $E\ge 5\times 10^{18}$ eV can be fitted at the same time only at the price of requiring very hard injection spectra for all nuclei ($\propto E^{-\gamma_g}$ with $\gamma_g=1\div 1.6$) and a maximum acceleration energy $E_{max}\le 5 Z\times 10^{18}$ eV (see \cite{Aloisio:2017qoo,Aloisio:2017ooo} and references therein). The need for hard spectra can be understood taking into account that the low energy tail of the flux of UHECR reproduces the injection power law. Therefore, taking $\gamma\ge 2$ cause the low energy part of the spectrum to be polluted by heavy nuclei thereby producing a disagreement with the light composition observed at low energy.

One should appreciate here the change of paradigm that these findings imply: while in the case of a pure proton composition it is needed to find sources and acceleration mechanisms able to energise CR protons up to energies larger than $10^{20}$ eV with steep injection ($\gamma_g\simeq 2.5\div 2.7$) \cite{Aloisio:2017qoo,Aloisio:2017ooo}, the Auger data require that the highest energy part of the spectrum ($E>5\times 10^{18}$ eV) has a flat injection ($\gamma_g\simeq 1.0\div 1.6$) being dominated by heavy nuclei with maximum energy not exceeding a few $\times$ $Z\times 10^{18}$ eV  \cite{Aloisio:2017qoo,Aloisio:2017ooo,Aloisio:2013hya}. 

By accepting the new paradigm, it follows that the Auger spectrum at energies below $5\times 10^{18}$ eV requires an additional component that, composed by protons and helium nuclei, could be, in principle, of galactic or extra-galactic origin \cite{Aloisio:2017qoo,Aloisio:2013hya}. Nevertheless, the anisotropy expected for a galactic light component extending up to $10^{18}$ eV exceeds by more than one order of magnitude the upper limit measured by Auger \cite{Abreu:2012ybu}. This observation, just restricting the analysis to Auger data, would constrain the transition between galactic and extra-galactic CR at energies below $10^{18}$ eV \cite{Aloisio:2017qoo,Aloisio:2012ba}, thus excluding a galactic component at the highest energies. 

In order to reproduce the Auger observations, the additional (light and extra-galactic) contribution to the flux at energies below $5\times 10^{18}$ eV should exhibit a steep power law injection at the source with $\gamma_g\simeq 2.6\div 2.7$ and a maximum acceleration energy not exceeding a few $\times$ $10^{18}$ eV, as for the heavier component  (see \cite{Aloisio:2017qoo,Aloisio:2017ooo} and references therein). The possible origin of this radiation can be modelled essentially in two ways: (i) assuming the presence of different classes of sources: one injecting also heavy nuclei with hard spectrum and the other only proton and helium nuclei with soft spectrum or (ii) identifying a peculiar class of sources that could provide at the same time a steep light component and a flat heavy one (see \cite{Aloisio:2017qoo,Aloisio:2017ooo} and references therein). The second approach is based on a specific hypothesis on the sources that should be surrounded by an intense radiation field that, through photo-disintegration of heavy nuclei in the source neighbourhood, can provide a light component of (secondary) protons with a steep spectrum together with a hard and heavier component. In figure \ref{fig1} we plot the flux and mass composition as observed by Auger and as reproduced theoretically assuming two classes of extragalactic sources with different injection characteristics as discussed above.

\section{Secondary neutrinos and gamma rays} 
\label{secondaries} 

The propagation of UHECR in the intergalactic space, through the interactions with CMB and EBL, gives rise to the production of several unstable particles that in turn produce high energy photons, electrons/positrons and neutrinos. The possible detection of these signal carriers, realised already at the end of sixties, is extremely important to constrain models for UHECR sources, composition and the details of propagation (see \cite{Aloisio:2017qoo,Aloisio:2017ooo} and references therein). In the following we will separately discuss the case of neutrinos and gamma rays.

There are two processes by which neutrinos can be emitted in the propagation of UHECR: (i) the decay of charged pions, produced by photo-pion production, $\pi^{\pm}\to \mu^{\pm}+\nu_{\mu}(\bar{\nu}_{\mu})$ and the subsequent muon decay $\mu^{\pm}\to e^{\pm}+\bar{\nu}_{\mu}(\nu_{\mu})+\nu_e(\bar{\nu}_e)$; (ii) the beta-decay of neutrons and nuclei produced by photo-disintegration: $n\to p+e^{-}+\bar{\nu}_e$, $(A,Z)\to (A,Z-1)+e^{+}+\nu_e$, or $(A,Z)\to (A,Z+1)+e^{-}+\bar{\nu}_e$. These processes produce neutrinos in different energy ranges: in the former the energy of each neutrino is around a few percent of that of the parent nucleon, whereas in the latter it is less than one part per thousand (in the case of neutron decay, larger for certain unstable nuclei). This means that in the interaction with CMB photons, which has a threshold Lorentz factor of about $\Gamma\ge 10^{10}$, neutrinos are produced with energies of the order of $10^{18}$~eV and $10^{16}$~eV respectively. Interactions with EBL photons contribute, with a lower probability than CMB photons, to the production of neutrinos with energies of the order of $10^{15}$~eV in the case of photo-pion production and $10^{14}$~eV in the case of neutron decay (see \cite{Aloisio:2015ega} and reference therein).   

\begin{figure}[!h]
\centering
\includegraphics[scale=.25]{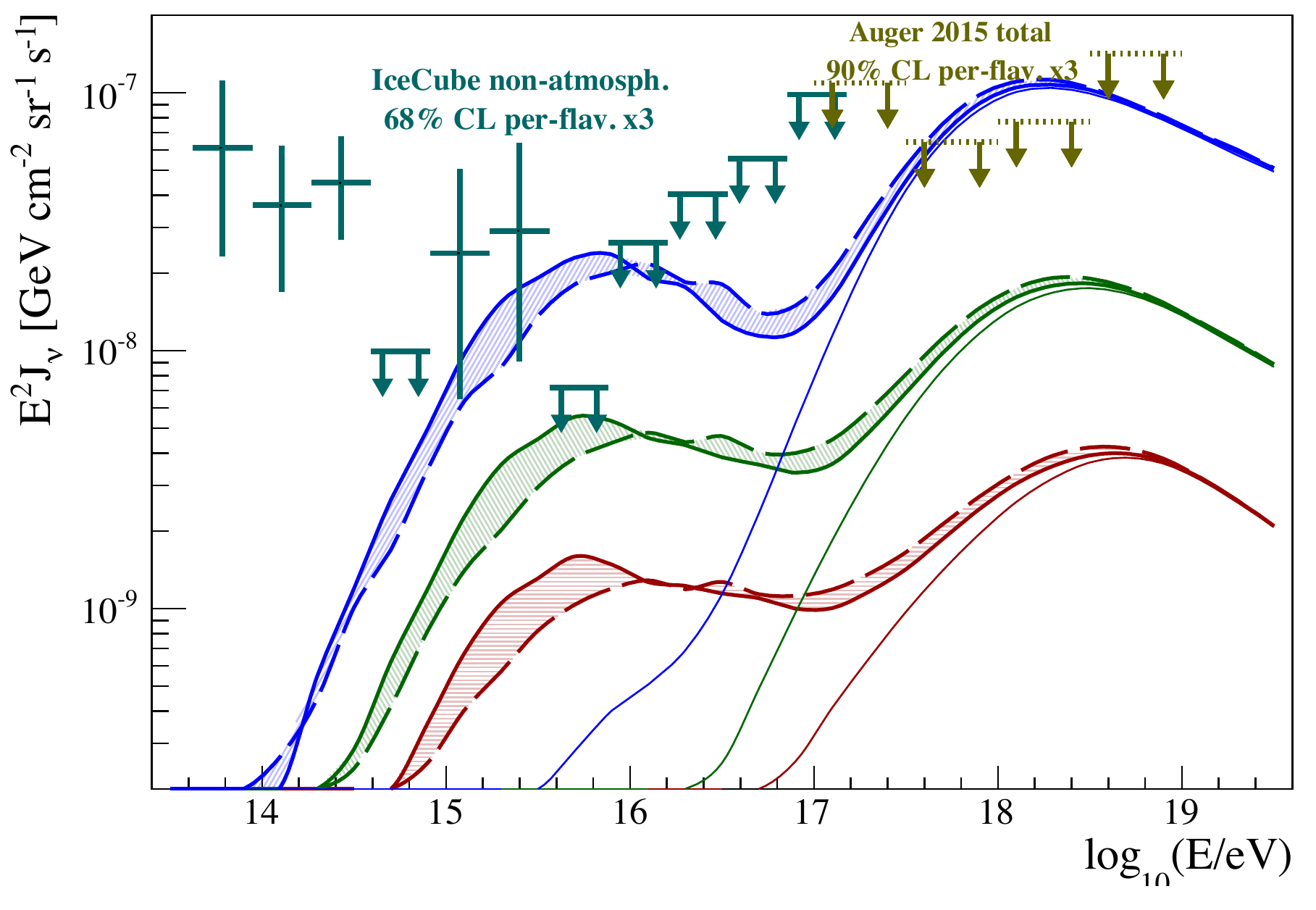}
\includegraphics[scale=.25]{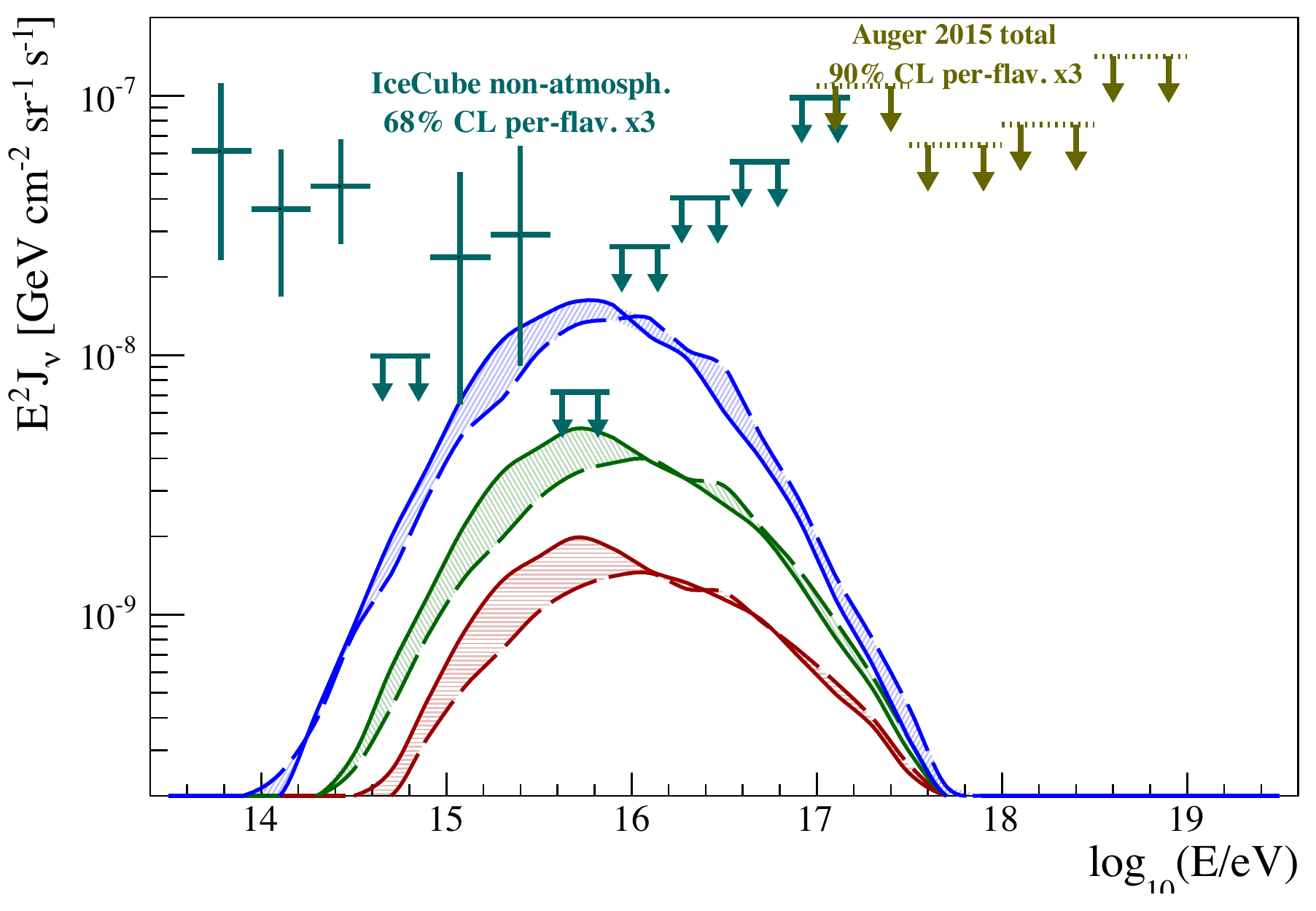}
\includegraphics[scale=.42]{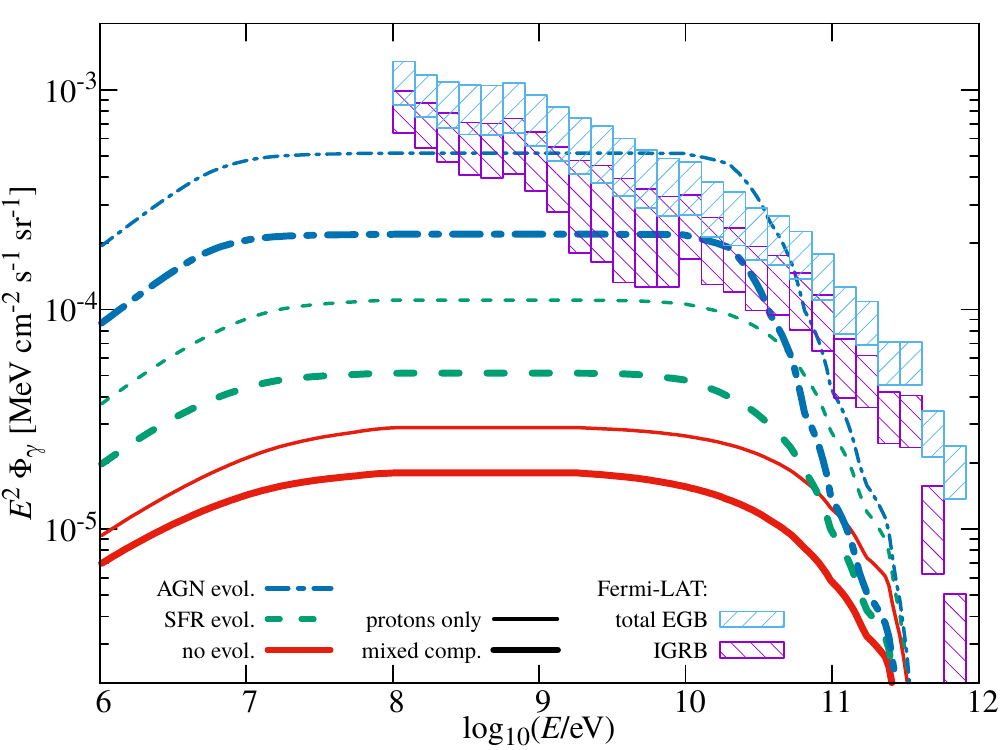}
\caption{ [Left panel] Fluxes of neutrinos in the case of a pure proton composition. The three different fluxes correspond to different assumptions on the cosmological evolution of sources (from bottom to top): no evolution (red), star formation rate evolution (green) and active galactic nuclei evolution (blue), coloured bands show the uncertainties due to the EBL model (see \cite{Aloisio:2017qoo,Aloisio:2017ooo} and references therein). Thin solid lines are neutrino fluxes obtained taking into account the sole CMB field. [Central Panel] Neutrino fluxes in the case of mixed composition, as shown in figure \ref{fig1}, with the same color code of left panel. Experimental points are the observations of IceCube on extra-terrestrial neutrinos \cite{Aartsen:2015awa} and the Auger limits on neutrino fluxes \cite{Abreu:2013zbq}. Figures taken from \cite{Aloisio:2015ega}. [Right Panel] Diffuse extragalactic gamma ray background expected from UHECR propagation in the two cases of a pure proton composition (thin lines) and mixed composition (thick lines) compared with Fermi-LAT observations \cite{Ackermann:2014usa}.} 
\label{fig2}
\end{figure}

The flux of secondary neutrinos is very much sensitive to the composition of UHECR. In figure \ref{fig2} we plot the flux of cosmogenic neutrinos expected in the case of a pure proton composition (left panel) and in the case of mixed composition (central panel). Comparing the two panels of figure \ref{fig2} it is evident the huge impact of the composition on the expected neutrino flux: heavy nuclei provide a reduced flux of neutrinos because the photo-pion production process in this case is subdominant. 

The production of cosmogenic neutrinos is almost independent of the variations in sources' distribution because the overall universe, up to the maximum red-shift, could contribute to the flux. Typically, the maximum red-shift of astrophysical structures is expected around $z_{max}\simeq 10$, which is the redshift of the first stars (pop III). Once produced at these cosmological distances neutrinos travel toward the observer almost freely, except for the adiabatic energy losses and flavour oscillations, the opacity of the universe to neutrinos being relevant only at redshifts $z \gg 10$. This is an important point that makes neutrinos a viable probe not only of the mass composition of UHECR but also of the cosmological evolution of sources. In figure \ref{fig2} (left and central panels) three different hypothesis on the cosmological evolution of sources are taken into account: no cosmological evolution (red bands), evolution typical of the star formation rate (green band) and of active galactic nuclei (blue band). 
 
There is a solid consensus about the light composition of UHECR in the low energy part of the observed spectrum. This assures a flux of cosmogenic neutrinos in the PeV energy region, produced by the protons' photo-pion production on the EBL photons. Coloured bands in figure \ref{fig2} show the uncertainties connected with the EBL background (see \cite{Aloisio:2017qoo,Aloisio:2017ooo,Aloisio:2015ega} and references therein). Another important uncertainty in the expected neutrino flux comes from the contribution of UHECR sources at high red-shift. Given the energy losses suffered by UHE protons and nuclei, sources at red-shift larger than $z>1$ can be observed only in terms of cosmogenic neutrinos \cite{Aloisio:2017qoo,Aloisio:2015ega}. 

While neutrinos reach the observer without being absorbed, high energy photons and electrons/positrons colliding with astrophysical photon backgrounds (CMB and EBL) produce Electromagnetic Cascades (EMC) through the processes of pair production (PP, $\gamma+\gamma_{CMB,EBL}\to e^{+}+e^{-}$) and Inverse Compton Scattering (ICS, $e+\gamma_{CMB,EBL}\to \gamma + e$). While PP is characterised by a threshold the ICS process does not. From this simple observation follows that once a cascade is started by a primary photon/electron/positron it develops since the energy of photons produced by ICS are still above the PP threshold. The final output of the cascade, i.e. what is left behind when the cascade is completely developed, is a flux of low energy photons all with energies below the PP threshold. 

The two astrophysical backgrounds CMB and EBL against which the EMC develops are characterised by typical energies $\epsilon_{CMB}\simeq 10^{-3}$ eV and $\epsilon_{EBL}\simeq 1$ eV. Hence, the typical threshold energy scale for pair-production will be respectively\footnote{Numerical values quoted here should be intended as reference values being background photons distributed over energy and not monochromatic.} ${\mathcal E}_{CMB}=m_e^2/\epsilon_{CMB}=2.5\times 10^{14}$ eV and ${\mathcal E}_{EBL}=m_e^2/\epsilon_{EBL}=2.5\times 10^{11}$ eV. The radiation left behind by the cascade will be restricted to energies below ${\mathcal E}_{EBL}$ and will pile up into the extragalactic gamma ray background.

In figure \ref{fig2} (right panel) we plot the expected gamma ray background produced by the propagation of UHECR assuming a pure proton composition (thin lines) or a mixed composition (thick lines). The case of pure protons maximises the number of secondary photons produced and, therefore, it is more constrained by the extragalactic gamma rays background, as observed by Fermi-LAT \cite{Ackermann:2014usa}. In the right panel of figure \ref{fig2} we have also considered different hypothesis on the cosmological evolution of sources, as for neutrinos, apart from the case of no cosmological evolution (solid lines), we considered the evolution typical of star formation rate (dashed lines) and of active galactic nuclei (dot dashed lines). The latter being only marginally consistent with the observed extragalactic gamma rays background. 

\section{Conclusions} 
\label{conclusions} 

The most important physical task in the physics of UHECR is certainly a clear identification of the sources. As discussed above, a key observable is the mass composition as it fixes few fundamental characteristics of the sources ($\gamma_g$ and $E_{max}$) and it regulates the production of secondary cosmogenic neutrinos and gamma rays. A pure proton composition is, theoretically, a natural possibility. Proton is the most abundant element in the universe and several different astrophysical objects, at present and past cosmological epochs, could provide efficient acceleration even if it requires very high luminosities and maximum acceleration energies. The complexity of the scenario based on a composition with heavy nuclei disfavours astrophysical sources placed at high redshift because of the lacking of heavy elements. To reproduce observations, mixed composition requires sources with flat injection spectra for heavy nuclei, or needs particular dynamics in the source environment, such as photo-disintegration on strong local photon fields. A remarkable feature of the mixed composition scenario is the relative low maximum energy required at the source.

We conclude highlighting the two principal avenues on which the study of UHECR should develop in the near future. From one hand, as discussed above, a firm experimental determination of the mass composition is an unavoidable step forward in this field of research. On the other hand, the highest energy regime, typically energies $E \gtrsim 5\times 10^{19}$ eV, still remains less probed with not enough statistics to firmly detect possible anisotropies in the arrival directions and the exact shape of the suppression. Current technologies, such as Auger and (extended) TA, can reach, in several years, one order of magnitude more in the number of observed events at the highest energies, which seems not enough to firmly detect anisotropies, to probe new physics or to detect neutrinos. New technologies are needed and future space observatories, with improved photon detection techniques, like the proposal POEMMA (Probe of Extreme Multi-Messenger Astrophysics) \cite{Olinto:2017xbi}, promise a new era in the physics of UHECR, allowing the needed statistics to detect anisotropies, cosmogenic neutrinos and to probe new physics \cite{Aloisio:2017qoo,Aloisio:2017ooo,Olinto:2017xbi}.

\bibliography{uhecr.bib}

\begin{thebibliography}{18}

\bibitem{Aloisio:2017qoo}
R.~Aloisio, PTEP \textbf{2017}, 12A102 (2017), \texttt{1707.08471}

\bibitem{Aloisio:2017ooo}
R.~Aloisio, P.~Blasi, I.~De~Mitri, S.~Petrera, Springer Book  (2018),
  \texttt{1707.06147}

\bibitem{ThePierreAuger:2015rma}
A.~Aab et~al. (Pierre Auger), Nucl. Instrum. Meth. \textbf{A798}, 172 (2015),
  \texttt{1502.01323}

\bibitem{Tinyakov:2014lla}
P.~Tinyakov (Telescope Array), Nucl. Instrum. Meth. \textbf{A742}, 29 (2014)

\bibitem{Aab:2017tyv}
A.~Aab et~al. (Pierre Auger), Science \textbf{357}, 1266 (2017),
  \texttt{1709.07321}

\bibitem{Aab:2014kda}
A.~Aab et~al. (Pierre Auger), Phys. Rev. \textbf{D90}, 122005 (2014),
  \texttt{1409.4809}

\bibitem{Abbasi:2014sfa}
R.U. Abbasi et~al., Astropart. Phys. \textbf{64}, 49 (2015), \texttt{1408.1726}

\bibitem{TheTelescopeArray:2018dje}
\emph{{Auger and TA: Joint Contributions to the ICRC 2017}},
  \texttt{1801.01018}

\bibitem{Aloisio:2015lva}
R.~Aloisio, S.~Matarrese, A.V. Olinto, JCAP \textbf{1508}, 024 (2015),
  \texttt{1504.01319}

\bibitem{Aloisio:2015ega}
R.~Aloisio et al,
  JCAP \textbf{1510}, 006 (2015), \texttt{1505.04020}

\bibitem{Abreu:2013env}
P.~Abreu et~al. (Pierre Auger), JCAP \textbf{1302}, 026 (2013),
  \texttt{1301.6637}

\bibitem{Aloisio:2013hya}
R.~Aloisio, V.~Berezinsky, P.~Blasi, JCAP \textbf{1410}, 020 (2014),
  \texttt{1312.7459}

\bibitem{Abreu:2012ybu}
P.~Abreu et~al. (Pierre Auger), Astrophys. J. \textbf{762}, L13 (2012),
  \texttt{1212.3083}

\bibitem{Aloisio:2012ba}
R.~Aloisio, V.~Berezinsky, A.~Gazizov, Astropart. Phys. \textbf{39-40}, 129
  (2012), \texttt{1211.0494}

\bibitem{Aartsen:2015awa}
M.G. Aartsen et~al. (IceCube) (2015), \texttt{1510.05225}

\bibitem{Abreu:2013zbq}
P.~Abreu et~al. (Pierre Auger), Adv.High Energy Phys. \textbf{2013}, 708680
  (2013), \texttt{1304.1630}

\bibitem{Ackermann:2014usa}
M.~Ackermann et~al. (Fermi-LAT), Astrophys. J. \textbf{799}, 86 (2015),
  \texttt{1410.3696}

\bibitem{Olinto:2017xbi}
A.V. Olinto et~al., PoS \textbf{ICRC2017}, 542 (2018), \texttt{1708.07599}

\end{thebibliography}

\end{document}